\documentclass{aa}
\usepackage[varg]{txfonts}
\usepackage{graphicx}

\usepackage[colorlinks=true, linkcolor=blue, urlcolor=blue, citecolor=blue]{hyperref}

\usepackage{natbib}
\bibpunct{(}{)}{;}{a}{}{,} 

\begin{document}

\title{A quantitative assessment of the VO line list: Inaccuracies hamper high-resolution VO detections in exoplanet atmospheres}
\titlerunning{A quantitative assessment of the VO line list}
\author{
S. de Regt\inst{\ref{inst1}} \and 
A. Y. Kesseli\inst{\ref{inst1}} \and 
I. A. G. Snellen\inst{\ref{inst1}} \and
S. R. Merritt\inst{\ref{inst2}} \and
K. L. Chubb\inst{\ref{inst3},\ref{inst4}}
}
\institute{
Leiden Observatory, Leiden University, Postbus 9513, 2300 RA, Leiden, The Netherlands\\\email{regt@strw.leidenuniv.nl}\label{inst1} \and
Astrophysics Research Centre, School of Mathematics and Physics, Queen’s University Belfast, Belfast BT7 1NN, UK\label{inst2} \and
Centre for Exoplanet Science, University of St Andrews, North Haugh, St Andrews, UK\label{inst3} \and
Stellar Astrophysics Centre, Deparment of Physics and Astronomy, Aarhus University, Ny Munkegade 120, DK-8000 Aarhus C, Denmark\label{inst4}
}

\date{Received 17 November 2021 / Accepted 20 February 2022}

\abstract
{Metal hydrides and oxides are important species in hot-Jupiters since they can affect their energy budgets and the thermal structure of their atmospheres. One such species is vanadium-oxide (VO), which is prominent in stellar M-dwarf spectra. Evidence for VO has been found in the low-resolution transmission spectrum of WASP-121b, but this has not been confirmed at high resolution. It has been suggested that this is due to inaccuracies in its line list.}
{In this paper, we quantitatively evaluate the VO line list and assess whether inaccuracies are indeed the reason for the non-detections at high resolution in WASP-121b. Furthermore, we investigate whether the detectability can be improved by selecting only those lines associated with the most accurate quantum transitions.}
{A cross-correlation analysis was applied to archival HARPS (High Accuracy Radial velocity Planet Searcher) and CARMENES spectra of several M dwarfs. VO cross-correlation signals from the spectra were compared with those in which synthetic VO models were injected, providing an estimate of the ratio between the potential strength (in case of a perfect model) and the observed strength of the signal. This was repeated for the reduced model covering the most accurate quantum transitions. The findings were subsequently fed into injection and recovery tests of VO in a UVES (Ultraviolet and Visual Echelle Spectrograph) transmission spectrum of WASP-121b.}
{We find that inaccuracies cause cross-correlation signals from VO in M-dwarf spectra to be suppressed by about a factor $2.1$ and $1.1$ for the complete and reduced line lists, respectively, corresponding to a reduced observing efficiency of a factor $4.3$ and $1.2$. The reduced line list outperforms the complete line list in recovering the actual VO signal in the M-dwarf spectra by about a factor of $1.8$. Neither line list results in a VO detection in WASP-121b. Injection tests show that with the reduced efficiency of the line lists, the potential signal as seen at low resolution is not detectable in these data.}
{}

\keywords{molecular data -- opacity -- stars: low-mass -- planets and satellites: atmospheres -- planets and satellites: individual: WASP-121b -- methods: observational -- techniques: spectroscopic
}

\maketitle
\section{Introduction}
Ultra-hot Jupiters (UHJs) are a class of gas giant exoplanets that orbit their host stars at extremely close distances and thus they are highly irradiated. Their equilibrium temperatures exceed  $\textit{T}_\mathrm{eq}\gtrsim2200\ \mathrm{K}$, akin to low-mass stars \citep{Lothringer_2019}. One of the defining features of UHJs is the presence of a thermal inversion, caused by spectroscopically active species absorbing stellar radiation in the upper atmosphere.
In this way, a thermal inversion affects a planet's energy budget and the redistribution of absorbed stellar energy from day to nightside. \citet{Hubeny_2003} and \citet{Fortney_2008} suggested that UHJ atmospheres are hot enough to contain gaseous titanium oxide (TiO) and vanadium oxide (VO), as in low-mass stars. These metal oxides could possibly be responsible for the inversions strongly absorbing incoming UV and optical radiation at high altitudes and heating up the upper atmosphere. 

In the case of UHJ WASP-121b, the presence of a thermal inversion has been observed using emission features in secondary eclipse Hubble Space Telescope (HST) spectra \citep{Evans_2017, Mikal_Evans_2019, Mikal_Evans_2020}, which is in line with the inefficient heat transport inferred from Transiting Exoplanet Survey Satellite (TESS) phase curve photometry \citep{Daylan_2019, Bourrier_2020}. Low-resolution spectroscopy has provided tentative evidence for VO in WASP-121b's atmosphere \citep{Evans_2016, Evans_2018, Mikal_Evans_2019}. High-resolution Doppler-resolved spectroscopy \citep{Snellen_2010, Brogi_2012, Birkby_2013} has been used in an effort to confirm the low-resolution VO detection. However, \citet{Hoeijmakers_2020} analysed transmission spectra observed by the High Accuracy Radial velocity Planet Searcher (HARPS) to report a non-detection of VO. Furthermore, \citet{Merritt_2020} present a non-detection using transmission spectra observed with the Ultraviolet and Visual Echelle Spectrograph (UVES), but both \citet{Merritt_2020} and \citet{Hoeijmakers_2020} stress that their VO non-detections are inconclusive as a consequence of the inaccuracies of the state-of-the-art ExoMol \citep{ExoMol_2020} VO line list \citep{McKemmish_2016}. Indeed, \citet{Hoeijmakers_2020} argue that their detection of V I implies the presence of VO as their equilibrium chemistry calculations show that a significant amount of vanadium should exist as gaseous VO.

In this paper, we present a quantitative assessment of the ExoMol VO line list to evaluate the discrepancy between the low-resolution evidence and the high-resolution non-detections of this molecule in the atmosphere of exoplanet WASP-121b. The line list's cross-correlation performance is studied using high-resolution spectra of M dwarf stars in Sect. \ref{Section2}. In Sect. \ref{Section3}, the results of the line-list assessment are used to interpret the analysis of UVES transmission spectra of WASP-121b. Section \ref{Section4} discusses the overall results and summarises the conclusions.

\section{VO line-list assessment} \label{Section2}
The most recent and accurate VO line list was constructed by the ExoMol group \citep{McKemmish_2016} and it is available from the ExoMol library\footnote{\url{https://www.exomol.com/data/molecules/VO/51V-16O/}}. The line list consists of approximately $640\,000$ energy levels between which more than 277 million transitions are transcribed. The energy levels are determined using quantum chemistry calculations and then adjusted based on experimental data. The limited availability of experimental data results in poorly constrained energy levels, which subsequently translate into large wavelength uncertainties of spectral lines. As a result, model spectra made with the ExoMol line list are not an exact representation of the actual VO opacity at high resolution. The presented analysis uses the ExoMol VO line list with version number 20160726, where additional experimental data were used to further refine the A$^{\mathrm{4}}\Pi$, B$^{\mathrm{4}}\Pi$, and C$^{\mathrm{4}}\Sigma^-$ states from the initial publication \citetext{Laura McKemmish, \textit{priv.\ comm.}}. There is a Measured Active Rotational-Vibrational Energy Levels (MARVEL) project for VO currently nearing completion \citep{Bowesman}; however, a new ExoMol line list is now also in production for VO which includes hyperfine splitting \citetext{Jonathan Tenysson, \textit{priv.\ comm.}}. This line list will be updated in the future with the MARVEL-produced high-accuracy energy levels for high-resolution studies. As highlighted by works such as \citet{McKemmish_2017} and \citet{Tennyson_2016a}, computing ab initio line lists for transition metal diatomics is challenging. The hyperfine splittings present in VO make it particularly complex.

\begin{figure*}[ht]
    \centering
    \includegraphics[width=17cm]{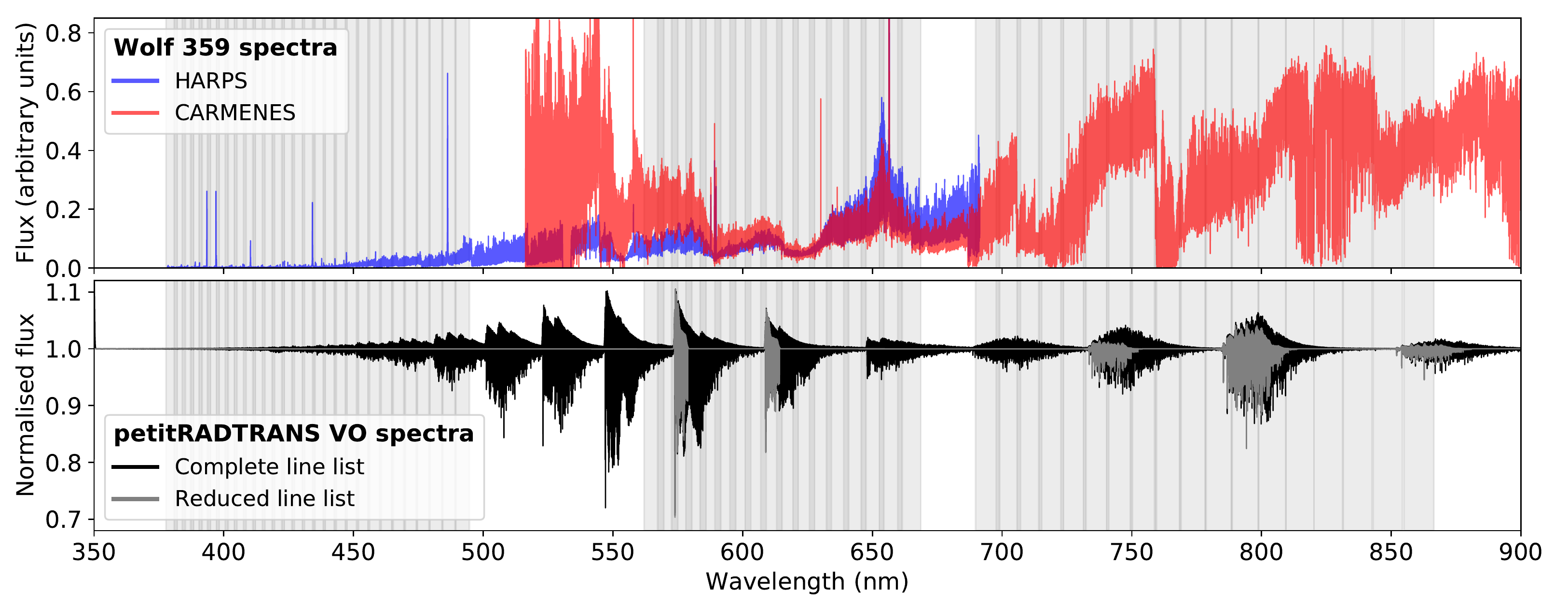}
    \caption{Comparison between the observed M-dwarf spectra and the petitRADTRANS VO model spectra. The top panel shows the HARPS and CARMENES spectra of Wolf 359 in blue and red, respectively. The bottom panel shows the normalised VO model spectra using all quantum transitions in black and using the specific transitions described in Sect. \ref{Section2.4} in grey. The grey vertical bands in both panels show the wavelength ranges of the UVES spectral orders.}
    \label{fig:spectra_comparison}
\end{figure*}

\subsection{M-dwarf spectra} \label{Section2.1}
The performance of the VO line list was quantified in a similar way as in \citet{McKemmish_2019} when updating the ExoMol TiO line list. We performed a cross-correlation with high-resolution spectra of M dwarfs as these stars have similar effective temperatures to UHJs and as VO is known to be an important opacity source \citep{Kirkpatrick_1999}. We utilised spectra observed with the HARPS and CARMENES\footnote{Calar Alto high-Resolution search for M dwarfs with Exoearths with Near-infrared and optical Echelle Spectrographs} spectrographs to cover a large wavelength range. The CARMENES optical arm has a spectral resolution of $\textit{R}\equiv\lambda/\Delta\lambda\sim94\,600$ and covers wavelengths between $520$-$960\ \mathrm{nm}$ \citep{Quirrenbach_2018}. The reduced spectra were retrieved from the CARMENES radial velocity survey \citep{Reiners_2018}. The HARPS spectrograph covers the wavelength range $380$-$690\ \mathrm{nm}$ with a spectral resolution of $\textit{R}\sim120\,000$ \citep{Mayor_2003}. The reduced data were retrieved from the ESO archive. The choice was made to focus on Wolf 359 as this bright star ($V=13.5$; \citealt{Landolt_1992}) is observed with both spectrographs. Additionally, Wolf 359 is a relatively cool ($\textit{T}_\mathrm{eff}=2800\ \mathrm{K}$; \citealt{Pavlenko_2006}), late-type star (M6.0; \citealt{Reiners_2018}) which prevents the thermal dissociation of VO and thus enhances its abundance. The observed spectra were shifted to the stellar rest frame by accounting for the barycentric and systemic velocity. The CARMENES orders were combined to produce a one-dimensional spectrum, which is already available for HARPS in the archive. The top panel of Fig. \ref{fig:spectra_comparison} shows the HARPS and CARMENES spectra of Wolf 359 in blue and red, respectively. The flux is in arbitrary units. The CARMENES spectrum has a low signal-to-noise near its blue edge, reflected by the large scatter in flux. The grey vertical bands show the wavelength ranges of the UVES spectral orders, obtained from the WASP-121b data used in Sect. \ref{Section3}.

\subsection{Model spectra}
We constructed model VO emission spectra using the radiative transfer code petitRADTRANS \citep{Molliere_2019}. This code can produce an emission or transmission spectrum at high or low resolution, given atmospheric parameters. petitRADTRANS uses opacity cross sections to compute the model spectra. To convert the ExoMol line list into these opacity data, we adopted the ExoCross code \citep{Yurchenko_2018} and followed the approach outlined in the petitRADTRANS documentation\footnote{\url{https://petitradtrans.readthedocs.io}}. We used a normalisation factor $\gamma=0.07\ \mathrm{cm}^{-1}$ for the pressure-broadening input \citep{Gharib_Nezhad_2019}. The computed opacities are available in the petitRADTRANS high-resolution opacity archive\footnote{\url{https://keeper.mpdl.mpg.de/d/e627411309ba4597a343}} as '\texttt{VO\_ExoMol\_McKemmish}'. Input for petitRADTRANS consists of a pressure-temperature ($\textit{PT}$) profile, the mass fractions of the requested species, the surface gravity ($\textit{g}$), and the mean molecular weight ($\mathrm{MMW}$). The $PT$ profile and $\mathrm{MMW}$ were retrieved from the Model Atmospheres in Radiative and Convective Scheme (MARCS) \citep{Gustafsson_2008}. Here, an effective temperature of $\textit{T}_\mathrm{eff}=2800\ \mathrm{K}$, a metallicity of $[\mathrm{Fe/H}]=0.0$, a surface gravity of $\textit{g}=10^{5.0}\ \mathrm{cm\ s}^{-2}$, and a microturbulence parameter of $\xi_\mathrm{t}=1\ \mathrm{km\ s}^{-1}$ were employed to model the Wolf 359 photosphere \citep{Pavlenko_2006}. Rayleigh scattering by H$_2$ and He was included as well as collision-induced absorption by H$_2$-H$_2$ and H$_2$-He pairs, and bound-free continuum absorption was included by H$^-$. The abundances of these species and VO were retrieved from the chemical equilibrium table which can be installed alongside petitRADTRANS \citep{Molliere_2017}. A C/O ratio of $0.62$ \citep{Nakajima_2016} was adopted to retrieve a mean VO volume-mixing ratio of $\mathrm{VMR}\sim1.4\cdot10^{-9}$. The PyAstronomy \citep{PyAstronomy} \texttt{rotBroad}-function was used to simulate rotational broadening with a projected velocity of $\textit{v}\sin i=2\ \mathrm{km\ s}^{-1}$ \citep{Reiners_2018}. Using a Gaussian filter, the VO model spectrum was broadened to the respective spectrograph's resolution. The bottom panel of Fig. \ref{fig:spectra_comparison} shows the normalised VO model spectrum in black. It is difficult to discern the VO absorption bands in the observed spectra because these include opacities from additional sources (mainly TiO; \citealt{Reiners_2018}).

\subsection{Cross-correlation} \label{Section2.3}
To assess the performance of the updated TiO line list, \citet{McKemmish_2019} cross-correlated their TiO model spectrum with both observed M-dwarf spectra and a synthetic PHOENIX spectrum, including all expected opacity sources, generated with the updated TiO line list. Rather than using fully synthetic spectra, we injected a Doppler-shifted VO model into the observed HARPS and CARMENES spectra. In this way, the injected VO signal and the actual signal were contained in the same spectrum, with similar opacities from other species (e.g. TiO) and similar noise properties. Hence, we can make a comparison between the optimal (injected) and the observed cross-correlation signal. Before injecting the model, we subtracted the corresponding blackbody profile to normalise the model spectrum. This normalised VO spectrum was Doppler-shifted with a radial velocity of $-25\ \mathrm{km\ s}^{-1}$ to avoid interference between the injected and true VO signals. Other radial velocities were also tested, but we found no significant differences in the results. After interpolating onto the observed spectrum's wavelength grid, the offset VO spectrum was multiplied into the observed spectrum, taking the respective instrumental resolutions with a Gaussian filter into account.

We applied a high-pass filter on both the observed and model spectra using $5\ \mathrm{\AA}$-wide Gaussian kernels to remove any broadband structures in the cross-correlation. The HARPS and CARMENES spectra were subsequently divided into the wavelength ranges of the UVES spectrograph's spectral orders. This division was carried out as we evaluated the VO line list's accuracy in the context of the non-detections in exoplanet WASP-121b. The wavelength ranges of the UVES orders were obtained from the data used in Sect. \ref{Section3}. Using the \texttt{crosscorrRV} routine from PyAstronomy \citep{PyAstronomy}, each of the orders was cross-correlated with the normalised VO model with velocities ranging from $-500$ to $+500\ \mathrm{km\ s}^{-1}$ in steps of $1\ \mathrm{km\ s}^{-1}$. The cross-correlation function (CCF) of each UVES-sized order was subsequently converted into a signal-to-noise function by dividing the entire CCF with the standard deviation outside of the injected and observed peaks ($|\textit{v}_\mathrm{rad}| > 100\ \mathrm{km\ s}^{-1}$). The values at $-25$ and $0\ \mathrm{km\ s}^{-1}$ were considered to be the injected and observed signal-to-noise ratios (S/Ns), respectively. 

\begin{figure}
    \resizebox{\hsize}{!}{\includegraphics{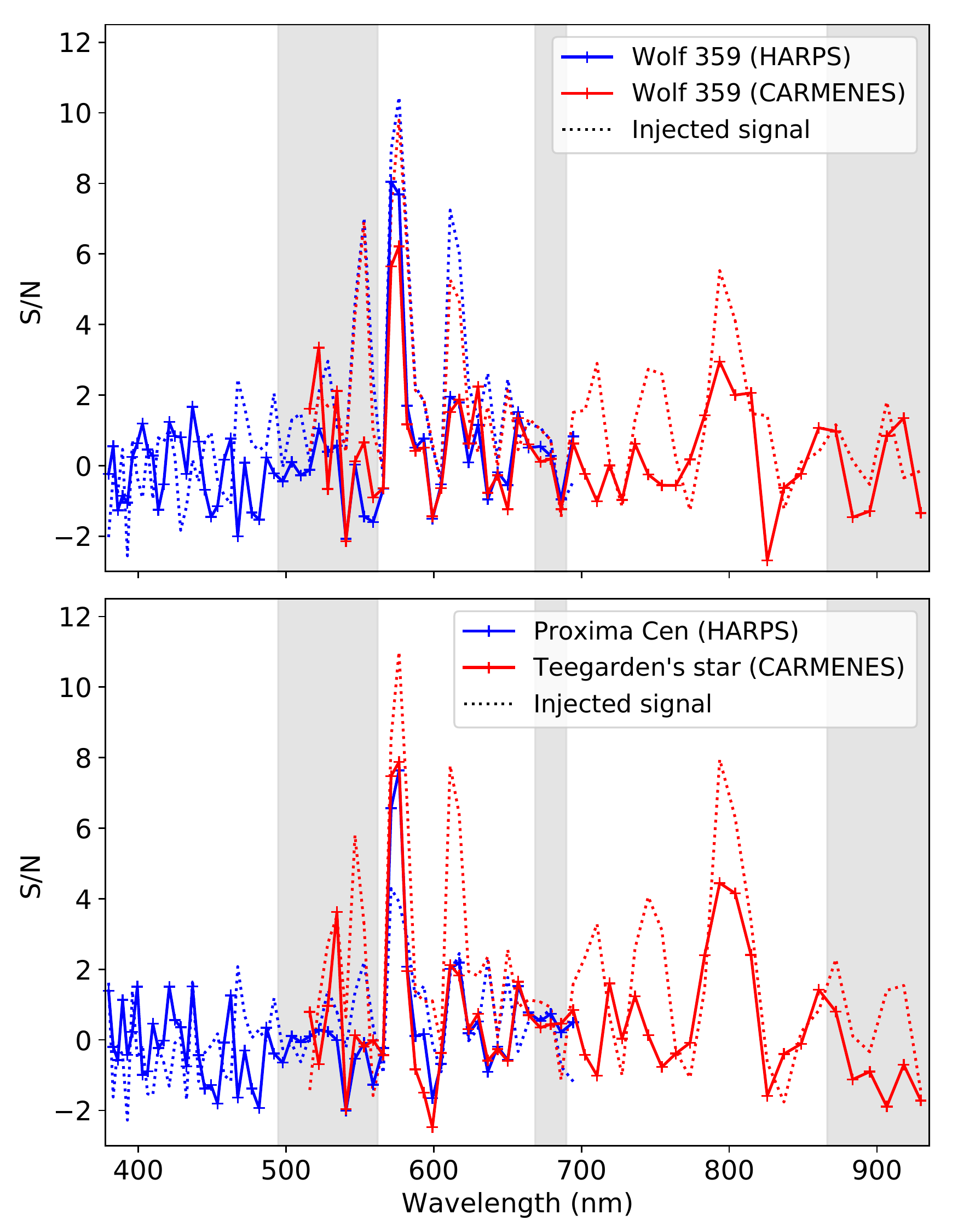}}
    \caption{Cross-correlation S/N between the ExoMol VO line list and M-dwarf spectra for each UVES-sized spectral order. The solid and dotted lines depict the observed and injected signals, respectively. The top panel shows the cross-correlation with the HARPS and CARMENES spectra of Wolf 359. The bottom panel shows the cross-correlation with the HARPS spectrum of Proxima Centauri and the CARMENES spectrum of Teegarden's star. The grey vertical bands are regions which are not covered by the UVES spectrograph.}
    \label{fig:SNR_vs_wave}
\end{figure}

Figure \ref{fig:SNR_vs_wave} shows the VO cross-correlation S/N for each UVES-sized order. Solid and dotted lines denote the observed and injected S/Ns, respectively. The top panel of Fig. \ref{fig:SNR_vs_wave} shows the cross-correlation with the HARPS (blue) and CARMENES (red) spectra of Wolf 359. The bottom panel displays the analysis of the HARPS spectrum of Proxima Centauri (M5.0) in blue and the CARMENES spectrum of Teegarden's star (M7.0) in red. Effective temperatures of $T_\mathrm{eff}=3000\ \mathrm{K}$ \citep{Ribas_2017} and $2700\ \mathrm{K}$ \citep{Kesseli_2019} were utilised to model the photospheres of Proxima Centauri and Teegarden's star, respectively. The grey vertical bands are areas that UVES does not cover, but they are included for completeness and future applications with other spectrographs. The similarity of the observed signals (solid lines) between the two spectrographs as well as the similarity between the different M dwarfs confirms that our analysis is broadly applicable for different spectrographs and spectral types and that it does not depend on an underlying noise structure. Peculiarly, Proxima Centauri's injected signal is lower than the observed signal around $\sim580\ \mathrm{nm}$, which is possibly a consequence of the adopted temperature leading to an inadequate VO abundance. In general, the observed and injected signals near the absorption bands of $\sim580$ and $\sim800\ \mathrm{nm}$ have comparable S/Ns. These absorption bands are the result of transitions of the electronic states $\mathrm{C}^4\Sigma^--\mathrm{X}^4\Sigma^-$ ($\sim580\ \mathrm{nm}$) and $\mathrm{B}^4\Pi-\mathrm{X}^4\Sigma^-$ ($\sim800\ \mathrm{nm}$). These transition energies are relatively well-refined with experimental data since the lowest vibrational quantum states are involved ($\nu=0$, \citealt{McKemmish_2016}). On the other hand, the bands at $\sim550$ and $\sim620\ \mathrm{nm}$ involve the less accurate, vibrationally excited states $\mathrm{C}^4\Sigma^-(\nu=1)$ and $\mathrm{X}^4\Sigma^-(\nu=1)$. As a consequence, the central wavelengths of these spectral lines are imprecise and the observed S/N is significantly lower than the injected, optimal S/N. Since the radial velocity shift of the injected signal is small, the low observed S/Ns at the less accurate band heads cannot be caused by TiO interference, for example, because that would affect the observed and injected signals equally. Rather, the injected model is cross-correlated with an exact copy, resulting in the optimal S/N. The S/N of the observed signal is lower because the VO template used in the cross-correlation is not a perfect duplicate of the actual VO absorption. Comparable results obtained with different injection velocities (e.g. $-50$, $+25\ \mathrm{km\ s}^{-1}$) support this interpretation.

\begin{figure}
    \resizebox{\hsize}{!}{\includegraphics{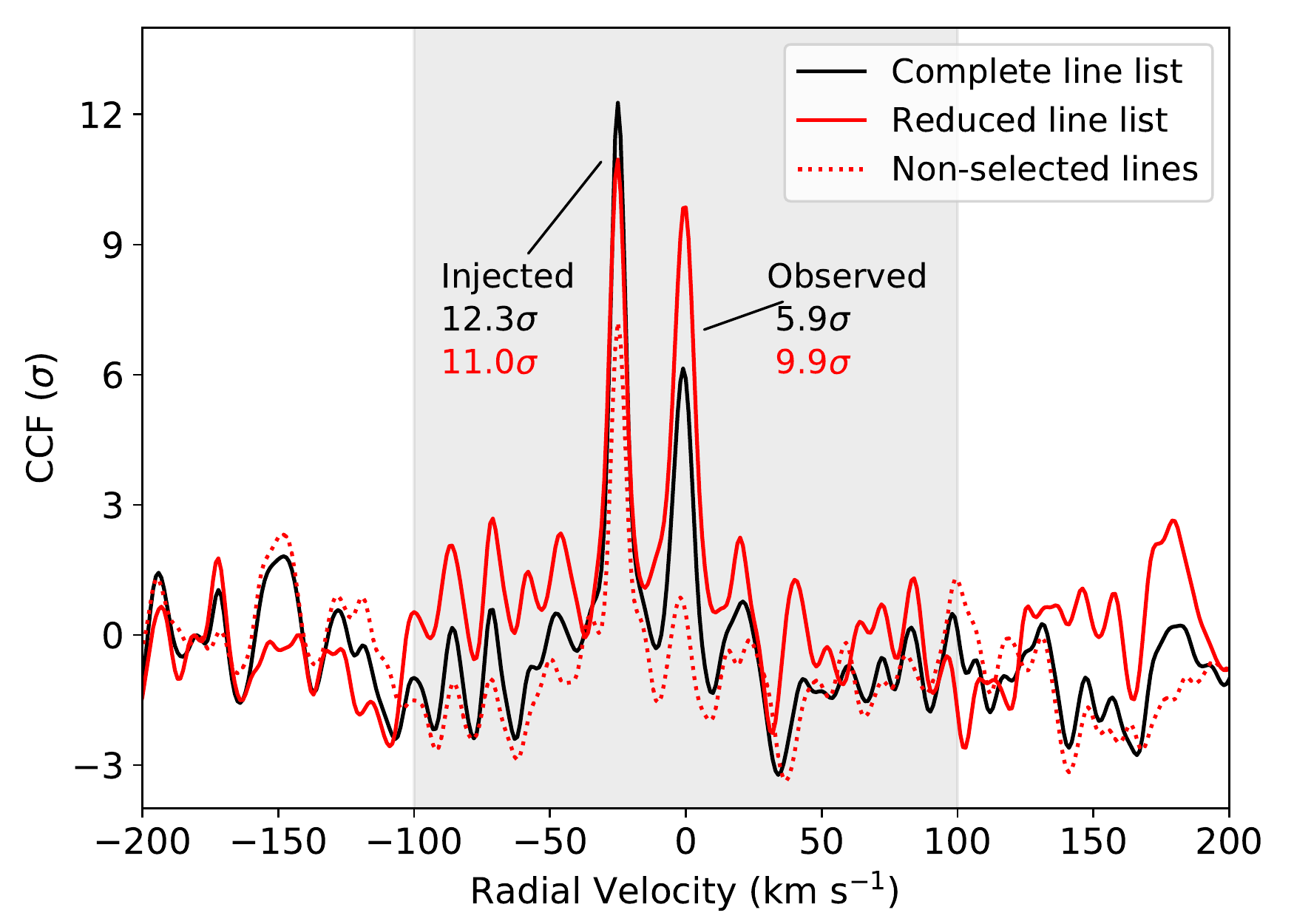}}
    \caption{Combined CCFs from every UVES-sized order after the VO injection. The black line shows the CCF using the complete VO line list. The solid and dotted red lines show the CCF using the reduced line list described in Sect. \ref{Section2.4} and the CCF using the non-selected transitions, respectively. The grey region depicts the excluded velocity range used to determine the noise level in the CCF. The peaks at $-25$ and $0\ \mathrm{km\ s}^{-1}$ are the injected and observed signals, respectively.}
    \label{fig:CCF}
\end{figure}

The CCFs of every UVES-sized order were summed together to obtain the total CCF. The orders bluewards of $578\ \mathrm{nm}$ utilised the CCFs with the HARPS spectrum of Wolf 359, while the redder orders used the CARMENES CCFs. The total CCF was converted into a signal-to-noise function by dividing with the standard deviation outside of $|\textit{v}_\mathrm{rad}| > 100\ \mathrm{km\ s}^{-1}$. The integrated CCF is shown in Fig. \ref{fig:CCF} as the black line. The total injected signal is detected at $12.3\sigma$ and the observed signal is detected at $5.9\sigma$, which makes the observed-to-injected signal ratio $(\mathrm{S/N})_\mathrm{obs} / (\mathrm{S/N})_\mathrm{inj}=0.48$. Hence, a cross-correlation analysis using the UVES-sized spectral orders and the ExoMol VO line list retrieves only $48\%$ of the potential signal from these M-dwarf spectra (reducing the signal by a factor $2.1$), requiring $2.1^2=4.3$ times more observing time. While this performance assessment is only correct if the utilised model parameters for Wolf 359 match the true parameters, we found that modified parameters (e.g. $T_\mathrm{eff}=2750$, $2850\ \mathrm{K}$ or $\log_{10} \textit{g}=4.5$, $5.5$) altered the observed-to-injected signal ratio by $\pm0.1$ at most. A different injection method was also tested, that is to say by the addition of the model spectrum in the observed spectra (instead of multiplication), but we did not find significant differences.

\subsection{Selection of accurate lines} \label{Section2.4}
Instead of using every quantum state in the VO line list, we also made models solely involving the well-constrained energy levels. Consequently, the inaccurate spectral lines are excluded and the model spectrum is a more accurate, but incomplete representation of the real VO absorption. The quantum states listed in Table \ref{tab1} were included in our reduced line list. These energy levels were refined with experimental data \citep[Table 3]{McKemmish_2016} and the spectroscopic model managed to fit the empirical energy levels relatively well with $\mathrm{RMS}(\omega)\lesssim 0.3\ \mathrm{cm}^{-1}$,  which corresponds to $\mathrm{RMS}(\textit{v}_\mathrm{rad})\lesssim 6\ \mathrm{km\ s}^{-1}$ at $700\ \mathrm{nm}$. Transitions from the X$^4\Sigma^-$ ground state to the excited B$^4\Pi$ or C$^4\Sigma^-$ states were included in the new reduced line list. This state selection decreases the number of transitions to 8008 between 669 unique upper states and 388 unique lower states. ExoCross was then run with a specific selection of transitions to generate the opacity data for petitRADTRANS. The computed opacity data are available in the petitRADTRANS archive\footnote{\url{https://keeper.mpdl.mpg.de/d/e627411309ba4597a343}} as '\texttt{VO\_ExoMol\_Specific\_Transitions}'. 

In the model spectrum generated with the complete line list, the large number of spectral lines caused a pseudo-continuum to form below the blackbody profile, decreasing the individual line depths. Since the reduced line list did not produce a similar pseudo-continuum, its spectral lines appeared deeper and this would have affected the cross-correlation if we had not performed a correction. The opacity cross sections of the reduced line list were subtracted from the complete line list, resulting in the opacities of the non-selected transitions. petitRADTRANS used the same configuration described in Sect. \ref{Section2.1} to generate a model spectrum of the non-selected transitions. Subsequently, we divided the complete model spectrum by the non-selected spectrum, generating a normalised model spectrum for the reduced line list including a correction for the line depths. As in Sect. \ref{Section2.3}, before cross-correlating, a high-pass filter was applied to the normalised model spectrum using a $5\ \mathrm{\AA}$-wide Gaussian kernel. The grey spectrum in the bottom panel of Fig. \ref{fig:spectra_comparison} shows the normalised model spectrum of the reduced line list. Many of the strongest spectral lines in the bandheads around $580$, $620$, $750$, $800$, and $870\ \mathrm{nm}$ are included in the state-selection model spectrum.

\begin{table}
\centering
\begin{tabular}{llll}
\hline
State & $\nu$ & $\Omega$ & J range \\
\hline
X$^{\mathrm{4}}\Sigma^-$ & 0 & $\pm$0.5 & 0.5 - 50.5 \\
                         &   & $\pm$1.5 & 1.5 - 50.5 \\
                         & 1 & $\pm$0.5 & 0.5 - 50.5 \\
                         &   & $\pm$1.5 & 1.5 - 50.5 \\
\hline
B$^{\mathrm{4}}\Pi$ & 0 & $-$0.5 & 13.5 - 47.5 \\
                    &   & $+$0.5 & 4.5 - 45.5 \\
                    &   & $\pm$1.5 & 7.5 - 40.5 \\
                    &   & $\pm$2.5 & 7.5 - 47.5 \\
                    & 1 & $-$0.5 & 5.5 - 24.5 \\
                    &   & $+$0.5 & 10.5 - 35.5 \\
                    &   & $\pm$1.5 & 9.5 - 31.5 \\
                    &   & $\pm$2.5 & 7.5 - 33.5 \\
\hline
C$^{\mathrm{4}}\Sigma^-$ & 0 & $\pm$0.5 & 0.5 - 41.5 \\
                         &   & $\pm$1.5 & 1.5 - 38.5 \\
\hline
\end{tabular}
\caption{Well-defined energy levels of VO \citep{McKemmish_2016} included in the reduced line list. The electronic state, vibrational quantum number $\nu$, the total electronic angular momentum $\Omega$, and the range of rotational quantum numbers J are listed.}
\label{tab1}
\end{table}

The spectrum derived from the reduced line list was cross-correlated with the spectra of Wolf 359 where the complete VO model was injected. The integrated CCF was obtained using the same method described in Sect. \ref{Section2.3} and is shown in Fig. \ref{fig:CCF} as a solid red line. The injected signal has an S/N of $11.0\sigma$, which is lower than that obtained with the complete line list ($12.3\sigma$). This is expected because the injected signal was recovered with substantially fewer spectral lines. On the other hand, the S/N of the observed signal has increased from $5.9\sigma$ to $9.9\sigma$, which confirms an improved accuracy by about a factor $1.8$ for the reduced line list. The observed-to-injected signal ratio using the reduced line list was calculated to be $(\mathrm{S/N})_\mathrm{obs}/(\mathrm{S/N})_\mathrm{inj}=0.90$. Hence, a cross-correlation analysis using the UVES-sized spectral orders and the reduced VO line list retrieves $90\%$ of the potential signal from these M-dwarf spectra (reducing the signal by a factor $1.1$), requiring $1.1^2=1.2$ times more observing time. Additionally, we performed a cross-correlation with the non-selected lines to confirm that most of the accurate lines were included in the reduced line list. The total CCF of the non-selected lines with the spectra of Wolf 359 is shown as the red dotted line in Fig. \ref{fig:CCF}. Since the observed signal measures at only $0.5\sigma$, we conclude that the non-selected transitions contribute to reducing the cross-correlation performance of the complete line list.

\section{Analysis of WASP-121b transmission spectra} \label{Section3}
The M dwarf analysis allowed us to determine the cross-correlation performance of the ExoMol VO line list, and to test the selection of specific transitions, increasing the VO detection significance. Subsequently, we aimed to confirm the presence of VO in the transmission spectrum of WASP-121b at high spectral resolution using the complete and reduced line list. Evidence for VO has been presented at low spectral resolution in the literature \citep{Evans_2016, Evans_2018, Mikal_Evans_2019}. WASP-121b orbits a bright F6V-type star ($\mathrm{V}=10.5$; \citealt{Hog_2000}) with a short orbital period of $1.27\ \mathrm{days}$ \citep{Delrez_2016}. The parameters of the WASP-121 system utilised in this paper are listed in Table \ref{tab2}. We used the same UVES transmission spectra previously presented by \citet{Gibson_2020}, \citet{Merritt_2020}, and \citet{Merritt_2021}. While \citet{Merritt_2020} used only the UVES red arm data for their non-detections of TiO and VO, our analysis used both the blue and red arms. The spectral resolution of the data was $R\sim80\,000$ for the blue arm and $R\sim110\,000$ for the red arm. A more detailed description of the data as well as an outline of the custom calibration pipeline can be found in \citet{Merritt_2020}. The pipeline places the spectra on a common wavelength grid and shifts the spectra to the stellar rest frame by correcting for the barycentric and systemic velocity. The stellar reflex motion is not accounted for because the stellar velocity semi-amplitude ($\textit{K}_\star={0.181}\ \mathrm{km\ s}^{-1}$; \citealt{Delrez_2016}) is significantly smaller than the resolution of the UVES spectrograph ($\sim{2.7}\ \mathrm{km\ s}^{-1}$ at $R\sim110\,000$).

\begin{table}
\centering
\begin{tabular}{ll}
\hline
\textbf{WASP-121} & \\ \hline
$\textit{M}_\star\ (\textit{M}_\odot)$ & $1.353_{-0.079}^{+0.080}$ $^\mathrm{a}$ \\
$\textit{R}_\star\ (\textit{R}_\odot)$ & $1.458\pm0.080$ $^\mathrm{a}$ \\
$\textit{v}_\mathrm{sys}\ (\mathrm{km\ s}^{-1})$ & $38.36\pm0.43$ $^\mathrm{b}$ \\
Limb-darkening: $\textit{c}_1$ & $0.395\pm0.003$ $^\mathrm{c}$ \\
Limb-darkening: $\textit{c}_2$ & $0.141\pm0.004$ $^\mathrm{c}$ \\
& \\
\textbf{WASP-121b} & \\ \hline
$\textit{T}_0\ (\mathrm{BJD}_{(\textit{TDB})})$ & $2457599.551478$ \\
                                                & $\pm0.000049$ $^\mathrm{d}$ \\
$\textit{P}\ (\mathrm{days})$ & $1.2749247646$ \\
                              & $\pm0.0000000714$ $^\mathrm{d}$ \\
$\textit{a}/\textit{R}_\star$ & $3.86\pm0.02$ $^\mathrm{e}$ \\
$\textit{R}_\mathrm{p}/\textit{R}_\star$ & $0.1218\pm0.0004$ $^\mathrm{e}$ \\
$\textit{M}_\mathrm{p}\ (M_\mathrm{Jup})$ & $1.183_{-0.062}^{+0.064}$ $^\mathrm{a}$ \\
$\textit{K}_\mathrm{p}\ (\mathrm{km\ s}^{-1})$ & $\sim217$ $^\mathrm{f}$ \\
$\textit{v}_\mathrm{eq}\ (\mathrm{km\ s}^{-1})$ & $7.1$ $^\mathrm{f}$ \\
& \\
\textbf{petitRADTRANS models} & \\
$\textit{P}_\mathrm{cloud}\ (\mathrm{mbar})$ & $20$ $^\mathrm{e}$ \\
$\textit{T}\ (\mathrm{K})$ & $1500$, $2000$, $2500$ and $3000$ \\
$\log_{10}\mathrm{VMR}\ [\mathrm{VO}]$ & $-13$, $-12$, $-11$, $-10$, $-9$, $-8$, \\
& $-7$, $-6$, $-5$, $-4$, $-3$ and $-6.6$ $^\mathrm{e}$ \\
\hline
\end{tabular}
\caption{System parameters of WASP-121b and its host star as used in this paper. Values marked with (a) are adopted from \citet{Delrez_2016}; (b) from \citet{Gaia_Collaboration_2018}; (c) from \citet{Wilson_2021}; (d) from \citet{Sing_2019}; and (e) from \citet{Evans_2018}; and (f) were derived from the provided parameters. The equatorial rotation velocity $\textit{v}_\mathrm{eq}$ assumes a tidally locked planet.}
\label{tab2}
\end{table}

The variation of the blaze function was removed with the same method as \citet{Merritt_2020}. Since the blaze correction was found to be unstable at the edges of each spectral order, we removed the first 600 pixels and last 60 pixels of each order in the blue arm ($22\%$ of pixels; \citealt{Gibson_2020}), as well as the first 500 pixels from the red arm's orders ($12\%$ of pixels; \citealt{Merritt_2020}). We determined common features between the 134 spectra with principal-component analysis (PCA). Injection tests enabled us to determine that subtracting the first eight PCs removed the (quasi)-static stellar or telluric features sufficiently.

\subsection{Model transmission spectra} \label{Section3.2}
The petitRADTRANS radiative transfer code \citep{Molliere_2019} was used to make model transmission spectra of VO in WASP-121b's atmosphere. The utilised model parameters are listed in Table \ref{tab2}. As in \citet{Merritt_2020}, we used the same temperatures and cloud deck, and we assumed isothermal atmospheres with 100 atmospheric layers. Twelve constant VO volume-mixing ratios were tested, including $\mathrm{VMR\ [VO]}=10^{-6.6}$ as reported by \citet{Evans_2018}. Rayleigh scattering from $\mathrm{H}_2$ and continuum absorption by $\mathrm{H}^-$ were included. The necessary abundances of $\mathrm{H}_2$, $\mathrm{H}$ and $\mathrm{e}^-$ were produced by the petitRADTRANS chemical equilibrium table ($\mathrm{VMR}\ [\mathrm{H}_2]=0.2$, $\mathrm{VMR}\ [\mathrm{H}]=0.1$, $\mathrm{VMR}\ [\mathrm{e}^-]=2.1\cdot10^{-10}$) and we used the volume-mixing ratio of \citet{Evans_2018} for $\mathrm{H}^-$ ($\mathrm{VMR}\ [\mathrm{H}^-]=5\cdot10^{-10}$). A mean molecular weight of $2.33$ was used and the model spectra were rotationally broadened following the method of \citet{Brogi_2016} with $\textit{v}_\mathrm{eq}=7.1\ \mathrm{km\ s}^{-1}$, which was inferred by assuming tidal locking. The instrumental broadening was also accounted for with a Gaussian filter set by the resolution of the respective UVES arms. Using the same procedure outlined in Sect. \ref{Section2.4}, we performed a pseudo-continuum correction to obtain transmission spectra of the reduced line list with the appropriate line depths. Figure \ref{fig:model_transmission_spectra} in the appendix shows a comparison between the transmission spectra with different scale heights and temperatures. Before performing the cross-correlation analysis, the low-frequency structure was removed from each model transmission spectrum by dividing a smoothed spectrum, which was obtained by convolution with a $5\ \mathrm{\AA}$-wide Gaussian kernel. This normalisation of the model spectra was performed because the low-frequency structure in the observed data was removed by the detrending.

\subsection{Cross-correlation} \label{Section3.3}
Using the weighted cross-correlation function from Eq. 11 in \citet{Brogi_2019}, each spectrum of each order was cross-correlated with the model transmission spectra with radial velocities ranging from $-600$ to $+600\ \mathrm{km\ s}^{-1}$ in steps of $0.5\ \mathrm{km\ s}^{-1}$. The out-of-transit spectra do not contain a planetary signal and the signal is weaker for the frames during ingress and egress. The varying signal strength was accounted for by weighting the CCFs with a PyTransit \citep{Parviainen_2015} model light curve using the equations of \citet{Mandel_2002} with limb-darkening coefficients $\textit{c}_1=0.395$ and $\textit{c}_2=0.141$ \citep{Wilson_2021}. We note that VO is absent in the stellar atmosphere due to WASP-121's high effective temperature ($6459\pm140\ \mathrm{K}$; \citealt{Delrez_2016}), and thus we did not encounter the Rossiter-McLaughlin effect or centre-to-limb variation. Following the integration steps described in \citet{Merritt_2020}, a $\textit{K}_\mathrm{p}$-$\textit{v}_\mathrm{sys}$ map of cross-correlation coefficients was constructed for each evaluated model. A coefficient was calculated for planet velocities $\textit{K}_\mathrm{p}$ ranging from $-400$ to $+400\ \mathrm{km\ s}^{-1}$ in steps of $1\ \mathrm{km\ s}^{-1}$ and systemic velocities $\textit{v}_\mathrm{sys}$ ranging from $-200$ to $+200\ \mathrm{km\ s}^{-1}$ in steps of $1\ \mathrm{km\ s}^{-1}$. The $\textit{K}_\mathrm{p}$-$\textit{v}_\mathrm{sys}$ maps were converted into detection S/Ns by dividing with the standard deviation of two rectangles outside of the expected peak ($\textit{K}_\mathrm{p}$ from $+100$ to $+300\ \mathrm{km\ s}^{-1}$ and $\textit{v}_\mathrm{sys}$ from $-200$ to $-50$ or from $+50$ to $+200\ \mathrm{km\ s}^{-1}$). We set our detection threshold at $\mathrm{4}\sigma$, since we made a simple noise estimation and noise fluctuations of the $\textit{K}_\mathrm{p}$-$\textit{v}_\mathrm{sys}$ maps could therefore be misinterpreted as detections with a lower threshold \citep{Cabot_2018}. 

\begin{figure*}[ht]
    \centering
    \includegraphics[width=17cm]{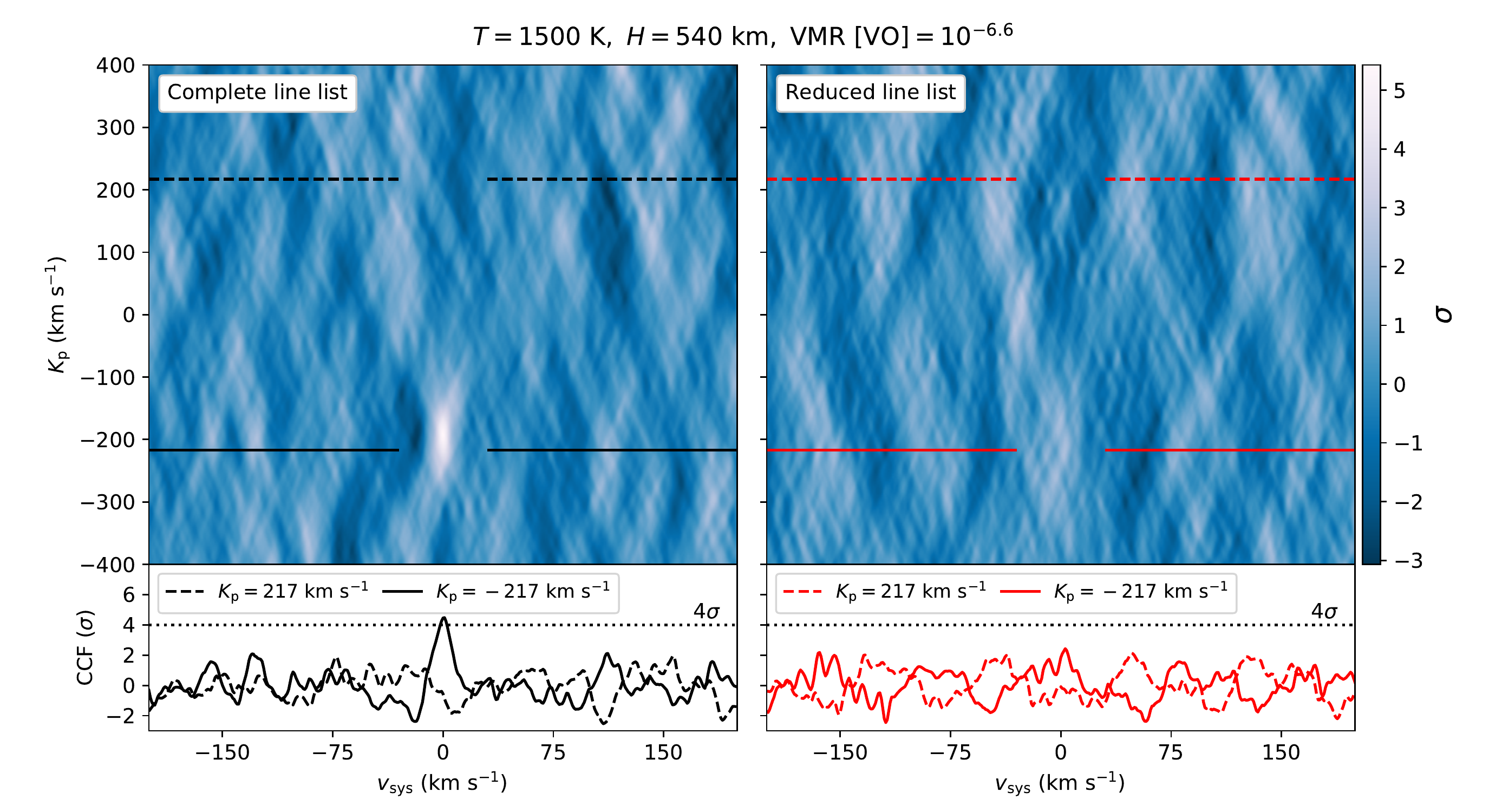}
    \caption{Cross-correlation analysis of WASP-121b's UVES transmission spectra, using the model parameters reported by \citet{Evans_2018}. The left panels show the analysis with the complete line list and the right panels use the reduced line list described in Sect. \ref{Section2.4}. The top panels show the $\textit{K}_\mathrm{p}$-$\textit{v}_\mathrm{sys}$ maps and the bottom panels show horizontal slices of these maps at the expected planet velocity $\textit{K}_\mathrm{p}=217\ \mathrm{km\ s}^{-1}$ (dashed) and at the injection velocity $\textit{K}_\mathrm{p}=-217\ \mathrm{km\ s}^{-1}$ (solid).}
    \label{fig:SNR_maps}
\end{figure*}

\begin{figure*}
    \centering
    \includegraphics[width=17cm]{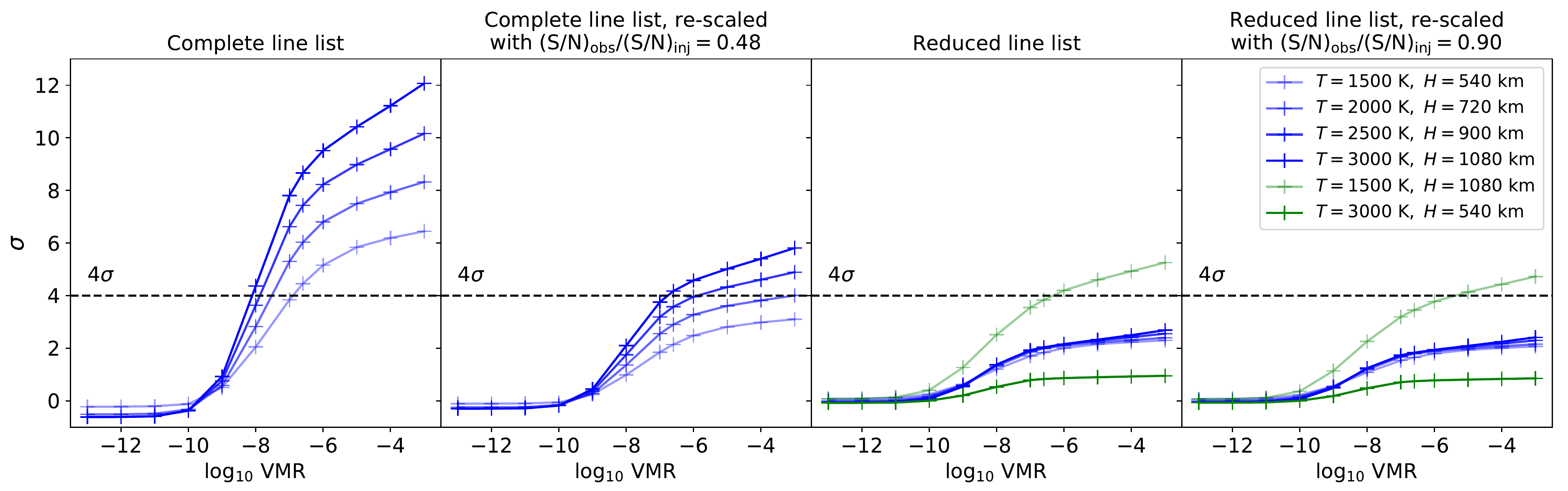}
    \caption{Detection significance of the injected signal versus $\log_\mathrm{10}\mathrm{VMR}$ for each model configuration. The left panel shows the recovered signal using all quantum transitions. The second panel shows the same signal multiplied by $0.48$ as found in Sect. \ref{Section2.1}. The third panel shows the recovered signal using specific quantum transitions and the fourth panel shows the same signal multiplied by $0.90$ as obtained in Sect. \ref{Section2.4}. The considered  temperatures are displayed as different shades of blue. The two rightmost panels also show the detection significances of models with $T=1500\ \mathrm{K}$, $H=1080\ \mathrm{km}$, and $T=3000\ \mathrm{K}$, $H=540\ \mathrm{km}$ in light green and dark green, respectively. The horizontal dashed lines at $4\sigma$ represent our detection thresholds.}
    \label{fig:SNR_vs_VMR}
\end{figure*}

\subsection{Results and injection tests} \label{Section3.4}
In agreement with \citet{Merritt_2020}, our cross-correlation analysis failed to retrieve a significant signal around the expected velocities ($\textit{K}_\mathrm{p}\sim217\ \mathrm{km\ s}^{-1}$ and $\textit{v}_\mathrm{sys}\sim0\ \mathrm{km\ s}^{-1}$) with any of the evaluated configurations of $\mathrm{VMR}$ and $\textit{T}$. Furthermore, our reduced line list did not yield a VO detection around the expected $\textit{K}_\mathrm{p}$ and $\textit{v}_\mathrm{sys}$ with any of the model spectra. Figure \ref{fig:SNR_maps} displays the $\textit{K}_\mathrm{p}$-$\textit{v}_\mathrm{sys}$ maps of both assessed line lists for the atmosphere reported by \citet{Evans_2018} ($\mathrm{VMR}=10^{-6.6}$, $\textit{T}=1500\ \mathrm{K}$, and a derived scale height of $\textit{H}=540\ \mathrm{km}$). The colourbar and the dashed CCFs in the bottom panels demonstrate that neither line list recovers a signal exceeding the $4\sigma$ detection threshold around the expected $\textit{K}_\mathrm{p}$ and $\textit{v}_\mathrm{sys}$.

Injection tests were carried out to determine whether a VO transmission signal could be detected with our retrieval method. The injection was performed by multiplying the UVES spectra with the complete model transmission spectra before applying the blaze correction. The varying signal strength was accounted for by scaling the models with the same PyTransit \citep{Parviainen_2015} light curve described in Sect. \ref{Section3.3}. The models were injected at $\textit{K}_\mathrm{p}=-217\ \mathrm{km\ s}^{-1}$ to avoid the enhancement by any undetected, real VO signal. Figure \ref{fig:SNR_maps} displays the $\textit{K}_\mathrm{p}$-$\textit{v}_\mathrm{sys}$ maps around the injected signal and the solid CCFs in the bottom panels show the horizontal slices at $\textit{K}_\mathrm{p}=-217\ \mathrm{km\ s}^{-1}$. The retrieval with the complete line list yields a detectable signal of $4.5\sigma$. However, we emphasise that this injection test assumes a perfect line list as the cross-correlation template is identical to the injected model. Section \ref{Section2.3} demonstrates that this assumption is incorrect and that we need to correct for the inaccuracies of the ExoMol VO line list. Similar to the M dwarf injection test of Sect. \ref{Section2.4}, the reduced line list recovers a lower injected signal from the UVES spectra as a consequence of the decreased number of lines.

Figure \ref{fig:SNR_vs_VMR} displays the detection significance of the injected signal as a function of the $\log_{10}$ VO volume-mixing ratio. The leftmost panel shows the recovered signal using the complete VO line list. As expected, higher temperatures (indicated by the dark blue lines) increase the atmospheric scale height, which in turn increases the detection significance. Our more extensive treatment of rotational broadening decreases the detection S/Ns compared to analogous models (e.g. $T=1500\ \mathrm{K}$ and $H=550\ \mathrm{km}$) in Fig. 6 in \citet{Merritt_2020}. The results of the injection tests are more similar to those of \citet{Merritt_2020} when the effects of rotation are disregarded. Many evaluated models exceed the $4\sigma$ detection limit and should therefore be detectable. However, Sect. \ref{Section2.3} demonstrates that the ExoMol VO line list is imperfect and only $48\%$ of an optimal, injected cross-correlation signal could be recovered from M dwarf Wolf 359's spectrum. We utilised the derived cross-correlation performance to simulate the reduction of the signal recovered from the WASP-121b data caused by the line list's inaccuracies. The second panel of Fig. \ref{fig:SNR_vs_VMR} shows the injected VO signal multiplied by $0.48$. The injected signal no longer exceeds the $4\sigma$ detection limit with the parameters reported by \citet{Evans_2018}. In fact, the re-scaled VO signal is not detectable for $T=1500$ and $2000\ \mathrm{K}$ with any of the evaluated abundances. Hence, it appears unlikely that the low-resolution VO detection by \citet{Evans_2018} ($T=1500\ \mathrm{K}$, $\mathrm{VMR}=10^{-6.6}$) can be confirmed with these UVES spectra and the current VO line list. Via interpolation, we find that abundances higher than $\mathrm{VMR}=10^{-6.8}$ exceed the detection threshold for $T=3000\ \mathrm{K}$. However, at this temperature thermal dissociation is expected to decrease the VO abundance to $\lesssim10^{-10}$ \citep{Merritt_2020}.

The third panel of Fig. \ref{fig:SNR_vs_VMR} shows the recovered signal using the reduced line list and the rightmost figure shows the same signal multiplied by the observed-to-injected signal ratio for the reduced line list (Sect. \ref{Section2.4}; $(\mathrm{S/N})_\mathrm{obs} / (\mathrm{S/N})_\mathrm{inj}=0.90$). In contrast with the complete line list, the cross-correlation signal is not increased two-fold by a doubling of the scale height ($H=540$ and $1080\ \mathrm{km}$) following an increase in the temperature ($T=1500$ and $3000\ \mathrm{K}$). We diagnosed that this discrepancy is an inherent property of our reduced line list. The light green line in Fig. \ref{fig:SNR_vs_VMR} displays the injected signal obtained with model transmission spectra where we artificially set the scale height to $H=1080\ \mathrm{km}$ (by adjusting the surface gravity in petitRADTRANS) for a temperature of $T=1500\ \mathrm{K}$. For these models, the detection significances show an approximate two-fold increase over the $H=540\ \mathrm{km}$ models, caused by the increased depths of the absorption lines. On the other hand, a higher temperature causes the lines selected in our reduced line list to generally become shallower as a consequence of rising collisional excitations. The relatively low-lying states in our reduced line list become less populated and the opacity of their transitions therefore decreases. The dark green line in Fig. \ref{fig:SNR_vs_VMR} displays models with a heightened temperature of $T=3000\ \mathrm{K}$ for a scale height of $H=540\ \mathrm{km}$. The injected signal is substantially decreased due to the temperature effect. The increase brought about by the scale height and the decrease caused by the temperature effect yield a negligible improvement in the detection significance retrieved by our reduced line list for a rising temperature. The cancellation is also found in the bottom panel of Fig. \ref{fig:model_transmission_spectra}, where the bands that are used in the reduced line list have similar depths for $T=1500$ and $3000\ \mathrm{K}$. The complete line list retrieves an increased cross-correlation signal because the weaker lines involving the excited states become deeper by an increased temperature (e.g. $\sim600\ \mathrm{nm}$ in Fig. \ref{fig:model_transmission_spectra}). From Fig. \ref{fig:SNR_vs_VMR} we deduce that the reduced line list could not detect VO even if we assumed it to be perfect. Consequently, the evaluated models are not detectable after accounting for the reduced line list's inaccuracies. Therefore, our reduced line list also appears unsuited for detecting a VO signal in the transmission spectra of WASP-121b.

\section{Discussion and conclusions} \label{Section4}
We have presented a quantitative assessment of the ExoMol VO line list using high-resolution HARPS and CARMENES spectra of M dwarfs. The injection of a Doppler-shifted petitRADTRANS model spectra of VO allowed us to compare the cross-correlation performance of the injected, optimal signal against the observed signal. The line list performs well around the absorption bands at $\sim580$ and $\sim800\ \mathrm{nm}$ due to the availability of experimental data used in refining the computed energy levels \citep{McKemmish_2016}. Future work could focus on updating the vibrationally excited levels to improve the quality of the VO line list. We find that a cross-correlation analysis recovers only $48\%$ of the potential signal. Furthermore, we made a reduced VO line list which only included the most accurate energy levels. While this line list is an incomplete representation of the actual VO opacity, it achieves a higher cross-correlation signal in M dwarfs by a factor of $1.8$ compared to the complete line list. The reduced line list manages to recover $90\%$ of the optimal signal. We have presented non-detections of VO in the UVES transmission spectrum of WASP-121b using the complete and reduced line list. After accounting for the line list's performance, injection tests showed that our retrieval method would likely not detect VO if it were present in the abundance reported by \citet{Evans_2018}. This analysis appears to confirm that the VO non-detections from \citet{Hoeijmakers_2020} and \citet{Merritt_2020} are indeed inconclusive due to the inaccuracies of the ExoMol VO line list.

Recently, the ExoMol group released an updated TiO line list \citep{McKemmish_2019}\footnote{This line list was updated in August 2021 to correct an error which had been identified in the MARVEL process for TiO; the latest version number is 20210825, as can be found in the ExoMol .def file, \url{https://www.exomol.com/db/TiO/48Ti-16O/Toto/48Ti-16O__Toto.def}} after it was shown that the previous line lists were insufficient to retrieve a TiO cross-correlation signal in M-dwarf spectra (\citealt{Hoeijmakers_2015}; \citealt{Nugroho_2017}). The MARVEL algorithm \citep{MARVEL} analyses collated experimental data of TiO to determine experimental-accuracy energy levels. The TiO energy levels which had been computed using quantum chemistry were subsequently replaced with the MARVEL-produced energies. The updated TiO line list showed an improved performance when cross-correlating with high-resolution M-dwarf spectra. This MARVEL-isation has not yet been applied to the VO line list, but we note that work is currently underway in the ExoMol group to produce a high-resolution line list for VO which makes use of MARVEL energy levels \citetext{Jonathan Tennyson, \textit{priv. comm.}}. We expect such an update to greatly improve the ability to find this species in exoplanet atmospheres.

\begin{acknowledgements}
The authors would like to thank the HARPS consortium and CARMENES consortium for making their data publicly available. This work was further based on data retrieved at the European Organisation for Astronomical Research in the Southern Hemisphere as part of ESO program 098.C-0547. This work was performed using the ALICE compute resources provided by Leiden University. This work made use of the following software packages that were not referenced in the main text: NumPy, SciPy, AstroPy, Matplotlib, iPython and Pandas (\cite{van_der_Walt_2011}; \cite{Virtanen_2020}; \cite{Astropy_2018}; \cite{Hunter_2007}; \cite{Perez_2007}; \cite{Pandas_2020}). I.S. and A.K. acknowledge funding from the European Research Council (ERC) under the European Union's Horizon 2020 research and innovation program under grant agreement No 694513. 
\end{acknowledgements}

\bibliographystyle{aa}
\bibliography{References.bib}

\begin{thebibliography}{57}
\expandafter\ifx\csname natexlab\endcsname\relax\def\natexlab#1{#1}\fi

\bibitem[{Astropy~Collaboration {et~al.}(2018)Astropy~Collaboration, Sipőcz,
  Günther, Lim, Crawford, Conseil, Shupe, Craig, Dencheva, \&
  et~al.}]{Astropy_2018}
Astropy~Collaboration, Price-Whelan, A.~M., Sipőcz, B.~M., Günther, H.~M.,
  {et~al.} 2018, The Astronomical Journal, 156, 123

\bibitem[{Birkby {et~al.}(2013)Birkby, de~Kok, Brogi, de~Mooij, Schwarz,
  Albrecht, \& Snellen}]{Birkby_2013}
Birkby, J.~L., de~Kok, R.~J., Brogi, M., {et~al.} 2013, Monthly Notices of the
  Royal Astronomical Society: Letters, 436, L35–L39

\bibitem[{Bourrier {et~al.}(2020)Bourrier, Kitzmann, Kuntzer, Nascimbeni,
  Lendl, Lavie, Hoeijmakers, Pino, Ehrenreich, Heng, \& et~al.}]{Bourrier_2020}
Bourrier, V., Kitzmann, D., Kuntzer, T., {et~al.} 2020, Astronomy \&
  Astrophysics, 637, A36

\bibitem[{Bowesman {et~al.}((to be submitted))Bowesman, Akbari, Hopkins,
  Yurchenko, \& Tennyson}]{Bowesman}
Bowesman, C., Akbari, H., Hopkins, S., Yurchenko, S., \& Tennyson, J. (to be
  submitted), Journal of Quantitative Spectroscopy and Radiative Transfer

\bibitem[{Brogi {et~al.}(2016)Brogi, Kok, Albrecht, Snellen, Birkby, \&
  Schwarz}]{Brogi_2016}
Brogi, M., Kok, R. J.~d., Albrecht, S., {et~al.} 2016, The Astrophysical
  Journal, 817, 106

\bibitem[{Brogi \& Line(2019)}]{Brogi_2019}
Brogi, M. \& Line, M.~R. 2019, The Astronomical Journal, 157, 114

\bibitem[{Brogi {et~al.}(2012)Brogi, Snellen, de~Kok, Albrecht, Birkby, \&
  de~Mooij}]{Brogi_2012}
Brogi, M., Snellen, I. A.~G., de~Kok, R.~J., {et~al.} 2012, Nature, 486,
  502–504

\bibitem[{Cabot {et~al.}(2018)Cabot, Madhusudhan, Hawker, \&
  Gandhi}]{Cabot_2018}
Cabot, S. H.~C., Madhusudhan, N., Hawker, G.~A., \& Gandhi, S. 2018, Monthly
  Notices of the Royal Astronomical Society, 482, 4422–4436

\bibitem[{{Czesla} {et~al.}(2019){Czesla}, {Schr{\"o}ter}, {Schneider},
  {Huber}, {Pfeifer}, {Andreasen}, \& {Zechmeister}}]{PyAstronomy}
{Czesla}, S., {Schr{\"o}ter}, S., {Schneider}, C.~P., {et~al.} 2019, {PyA:
  Python astronomy-related packages}

\bibitem[{Daylan {et~al.}(2021)Daylan, Günther, Mikal-Evans, Sing, Wong,
  Shporer, Niraula, de~Wit, Koll, Parmentier, Fetherolf, Kane, Ricker,
  Vanderspek, Seager, Winn, Jenkins, Caldwell, Charbonneau, Henze, Paegert,
  Rinehart, Rose, Sha, Quintana, \& Villasenor}]{Daylan_2019}
Daylan, T., Günther, M.~N., Mikal-Evans, T., {et~al.} 2021, TESS observations
  of the WASP-121 b phase curve

\bibitem[{{Delrez} {et~al.}(2016){Delrez}, {Santerne}, {Almenara}, {Anderson},
  {Collier-Cameron}, {D{\'\i}az}, {Gillon}, {Hellier}, {Jehin}, {Lendl},
  {Maxted}, {Neveu-VanMalle}, {Pepe}, {Pollacco}, {Queloz}, {S{\'e}gransan},
  {Smalley}, {Smith}, {Triaud}, {Udry}, {Van Grootel}, \& {West}}]{Delrez_2016}
{Delrez}, L., {Santerne}, A., {Almenara}, J.~M., {et~al.} 2016, Monthly Notices
  of the Royal Astronomical Society, 458, 4025

\bibitem[{Evans {et~al.}(2018)Evans, Sing, Goyal, Nikolov, Marley, Zahnle,
  Henry, Barstow, Alam, Sanz-Forcada, \& et~al.}]{Evans_2018}
Evans, T.~M., Sing, D.~K., Goyal, J.~M., {et~al.} 2018, The Astronomical
  Journal, 156, 283

\bibitem[{Evans {et~al.}(2017)Evans, Sing, Kataria, Goyal, Nikolov, Wakeford,
  Deming, Marley, Amundsen, Ballester, \& et~al.}]{Evans_2017}
Evans, T.~M., Sing, D.~K., Kataria, T., {et~al.} 2017, Nature, 548, 58–61

\bibitem[{Evans {et~al.}(2016)Evans, Sing, Wakeford, Nikolov, Ballester,
  Drummond, Kataria, Gibson, Amundsen, \& Spake}]{Evans_2016}
Evans, T.~M., Sing, D.~K., Wakeford, H.~R., {et~al.} 2016, The Astrophysical
  Journal, 822, L4

\bibitem[{Fortney {et~al.}(2008)Fortney, Lodders, Marley, \&
  Freedman}]{Fortney_2008}
Fortney, J., Lodders, K., Marley, M., \& Freedman, R. 2008, The Astrophysical
  Journal, 678, 1419–1435

\bibitem[{{Furtenbacher} \& {Cs{\'a}sz{\'a}r}(2012)}]{MARVEL}
{Furtenbacher}, T. \& {Cs{\'a}sz{\'a}r}, A.~G. 2012, \jqsrt, 113, 929

\bibitem[{{Gaia Collaboration}(2018)}]{Gaia_Collaboration_2018}
{Gaia Collaboration}. 2018, VizieR Online Data Catalog, I/345

\bibitem[{Gharib-Nezhad \& Line(2019)}]{Gharib_Nezhad_2019}
Gharib-Nezhad, E. \& Line, M.~R. 2019, The Astrophysical Journal, 872, 27

\bibitem[{Gibson {et~al.}(2020)Gibson, Merritt, Nugroho, Cubillos, de~Mooij,
  Mikal-Evans, Fossati, Lothringer, Nikolov, Sing, \& et~al.}]{Gibson_2020}
Gibson, N.~P., Merritt, S., Nugroho, S.~K., {et~al.} 2020, Monthly Notices of
  the Royal Astronomical Society, 493, 2215–2228

\bibitem[{Gustafsson {et~al.}(2008)Gustafsson, Edvardsson, Eriksson,
  Jørgensen, Nordlund, \& Plez}]{Gustafsson_2008}
Gustafsson, B., Edvardsson, B., Eriksson, K., {et~al.} 2008, Astronomy \&
  Astrophysics, 486, 951–970

\bibitem[{Hoeijmakers {et~al.}(2015)Hoeijmakers, de~Kok, Snellen, Brogi,
  Birkby, \& Schwarz}]{Hoeijmakers_2015}
Hoeijmakers, H.~J., de~Kok, R.~J., Snellen, I. A.~G., {et~al.} 2015, Astronomy
  \& Astrophysics, 575, A20

\bibitem[{Hoeijmakers {et~al.}(2020)Hoeijmakers, Seidel, Pino, Kitzmann,
  Sindel, Ehrenreich, Oza, Bourrier, Allart, Gebek, \&
  et~al.}]{Hoeijmakers_2020}
Hoeijmakers, H.~J., Seidel, J.~V., Pino, L., {et~al.} 2020, Astronomy \&
  Astrophysics, 641, A123

\bibitem[{{H{\o}g} {et~al.}(2000){H{\o}g}, {Fabricius}, {Makarov}, {Urban},
  {Corbin}, {Wycoff}, {Bastian}, {Schwekendiek}, \& {Wicenec}}]{Hog_2000}
{H{\o}g}, E., {Fabricius}, C., {Makarov}, V.~V., {et~al.} 2000, Astronomy \&
  Astrophysics, 355, L27

\bibitem[{Hubeny {et~al.}(2003)Hubeny, Burrows, \& Sudarsky}]{Hubeny_2003}
Hubeny, I., Burrows, A., \& Sudarsky, D. 2003, The Astrophysical Journal, 594,
  1011–1018

\bibitem[{{Hunter}(2007)}]{Hunter_2007}
{Hunter}, J.~D. 2007, Computing in Science and Engineering, 9, 90

\bibitem[{Kesseli {et~al.}(2019)Kesseli, Davy~Kirkpatrick, Fajardo-Acosta,
  Penny, Scott~Gaudi, Veyette, Boeshaar, Henderson, Cushing, Calchi-Novati, \&
  et~al.}]{Kesseli_2019}
Kesseli, A.~Y., Davy~Kirkpatrick, J., Fajardo-Acosta, S.~B., {et~al.} 2019, The
  Astronomical Journal, 157, 63

\bibitem[{Kirkpatrick {et~al.}(1999)Kirkpatrick, Reid, Liebert, Cutri, Nelson,
  Beichman, Dahn, Monet, Gizis, \& Skrutskie}]{Kirkpatrick_1999}
Kirkpatrick, J.~D., Reid, I.~N., Liebert, J., {et~al.} 1999, The Astrophysical
  Journal, 519, 802

\bibitem[{{Landolt}(1992)}]{Landolt_1992}
{Landolt}, A.~U. 1992, The Astronomical Journal, 104, 340

\bibitem[{Lothringer \& Barman(2019)}]{Lothringer_2019}
Lothringer, J.~D. \& Barman, T. 2019, The Astrophysical Journal, 876, 69

\bibitem[{Mandel \& Agol(2002)}]{Mandel_2002}
Mandel, K. \& Agol, E. 2002, The Astrophysical Journal, 580, L171–L175

\bibitem[{{Mayor} {et~al.}(2003){Mayor}, {Pepe}, {Queloz}, {Bouchy},
  {Rupprecht}, {Lo Curto}, {Avila}, {Benz}, {Bertaux}, {Bonfils}, {Dall},
  {Dekker}, {Delabre}, {Eckert}, {Fleury}, {Gilliotte}, {Gojak}, {Guzman},
  {Kohler}, {Lizon}, {Longinotti}, {Lovis}, {Megevand}, {Pasquini}, {Reyes},
  {Sivan}, {Sosnowska}, {Soto}, {Udry}, {van Kesteren}, {Weber}, \&
  {Weilenmann}}]{Mayor_2003}
{Mayor}, M., {Pepe}, F., {Queloz}, D., {et~al.} 2003, The Messenger, 114, 20

\bibitem[{McKemmish {et~al.}(2019)McKemmish, Masseron, Hoeijmakers,
  Pérez-Mesa, Grimm, Yurchenko, \& Tennyson}]{McKemmish_2019}
McKemmish, L.~K., Masseron, T., Hoeijmakers, H.~J., {et~al.} 2019, Monthly
  Notices of the Royal Astronomical Society, 488, 2836–2854

\bibitem[{McKemmish {et~al.}(2017)McKemmish, Masseron, Sheppard, Sandeman,
  Schofield, Furtenbacher, Császár, Tennyson, \&
  Sousa-Silva}]{McKemmish_2017}
McKemmish, L.~K., Masseron, T., Sheppard, S., {et~al.} 2017, The Astrophysical
  Journal Supplement Series, 228, 15

\bibitem[{McKemmish {et~al.}(2016)McKemmish, Yurchenko, \&
  Tennyson}]{McKemmish_2016}
McKemmish, L.~K., Yurchenko, S.~N., \& Tennyson, J. 2016, Monthly Notices of
  the Royal Astronomical Society, 463, 771–793

\bibitem[{Merritt {et~al.}(2021)Merritt, Gibson, Nugroho, de~Mooij, Hooton,
  Lothringer, Matthews, Mikal-Evans, Nikolov, Sing, \& et~al.}]{Merritt_2021}
Merritt, S.~R., Gibson, N.~P., Nugroho, S.~K., {et~al.} 2021, Monthly Notices
  of the Royal Astronomical Society, 506, 3853–3871

\bibitem[{Merritt {et~al.}(2020)Merritt, Gibson, Nugroho, de~Mooij, Hooton,
  Matthews, McKemmish, Mikal-Evans, Nikolov, Sing, \& et~al.}]{Merritt_2020}
Merritt, S.~R., Gibson, N.~P., Nugroho, S.~K., {et~al.} 2020, Astronomy \&
  Astrophysics, 636, A117

\bibitem[{Mikal-Evans {et~al.}(2019)Mikal-Evans, Sing, Goyal, Drummond, Carter,
  Henry, Wakeford, Lewis, Marley, Tremblin, \& et~al.}]{Mikal_Evans_2019}
Mikal-Evans, T., Sing, D.~K., Goyal, J.~M., {et~al.} 2019, Monthly Notices of
  the Royal Astronomical Society, 488, 2222–2234

\bibitem[{Mikal-Evans {et~al.}(2020)Mikal-Evans, Sing, Kataria, Wakeford,
  Mayne, Lewis, Barstow, \& Spake}]{Mikal_Evans_2020}
Mikal-Evans, T., Sing, D.~K., Kataria, T., {et~al.} 2020, Monthly Notices of
  the Royal Astronomical Society, 496, 1638–1644

\bibitem[{Mollière {et~al.}(2017)Mollière, van Boekel, Bouwman, Henning,
  Lagage, \& Min}]{Molliere_2017}
Mollière, P., van Boekel, R., Bouwman, J., {et~al.} 2017, Astronomy \&
  Astrophysics, 600, A10

\bibitem[{Mollière {et~al.}(2019)Mollière, Wardenier, van Boekel, Henning,
  Molaverdikhani, \& Snellen}]{Molliere_2019}
Mollière, P., Wardenier, J.~P., van Boekel, R., {et~al.} 2019, Astronomy \&
  Astrophysics, 627, A67

\bibitem[{Nakajima \& Sorahana(2016)}]{Nakajima_2016}
Nakajima, T. \& Sorahana, S. 2016, The Astrophysical Journal, 830, 159

\bibitem[{Nugroho {et~al.}(2017)Nugroho, Kawahara, Masuda, Hirano, Kotani, \&
  Tajitsu}]{Nugroho_2017}
Nugroho, S.~K., Kawahara, H., Masuda, K., {et~al.} 2017, The Astronomical
  Journal, 154, 221

\bibitem[{Parviainen(2015)}]{Parviainen_2015}
Parviainen, H. 2015, Monthly Notices of the Royal Astronomical Society, 450,
  3233–3238

\bibitem[{Pavlenko {et~al.}(2006)Pavlenko, Jones, Lyubchik, Tennyson, \&
  Pinfield}]{Pavlenko_2006}
Pavlenko, Y.~V., Jones, H. R.~A., Lyubchik, Y., Tennyson, J., \& Pinfield,
  D.~J. 2006, Astronomy \& Astrophysics, 447, 709–717

\bibitem[{P\'erez \& Granger(2007)}]{Perez_2007}
P\'erez, F. \& Granger, B.~E. 2007, Computing in Science \& Engineering, 9, 21

\bibitem[{Quirrenbach {et~al.}(2018)Quirrenbach, Amado, Ribas, Reiners,
  Caballero, Seifert, Aceituno, Azzaro, Baroch, Barrado, Bauer, Becerril,
  Bèjar, Benítez, Brinkmöller, Guillén, Cifuentes, Colomé,
  Cortés-Contreras, Czesla, Dreizler, Frölich, Fuhrmeister,
  Galadí-Enríquez, Hernández, Peinado, Guenther, de~Guindos, Hagen, Hatzes,
  Hauschildt, Helmling, Henning, Herbort, Castaño, Herrero, Hintz, Jeffers,
  Johnson, de~Juan, Kaminski, Klahr, Kürster, Lafarga, Sairam, Lampón, Lara,
  Launhardt, del Fresno, López-Puertas, Luque, Mandel, Marfil, Martín,
  Martín-Ruiz, Mathar, Montes, Morales, Nagel, Nortmann, Nowak, Pallé,
  Passegger, Pavlov, Pedraz, Pérez-Medialdea, Perger, Rebolo, Reffert,
  Rodríguez, López, Rosich, Sabotta, Sadegi, Salz, Sánchez-López,
  Sanz-Forcada, Sarkis, Schäfer, Schiller, Schmitt, Schöfer, Schweitzer,
  Shulyak, Solano, Stahl, Pinto, Trifonov, Osorio, Yan, Zechmeister, Abellán,
  Abril, Alonso-Floriano, von Eiff, Anglada-Escudé, Anwand-Heerwart,
  Arroyo-Torres, Berdiñas, Bergondy, Blümcke, del Burgo, Cano, Carro,
  Cárdenas, Casal, Claret, Díez-Alonso, Doellinger, Dorda, Feiz, Fernández,
  Ferro, Gaisné, Gallardo, Gálvez-Ortiz, García-Piquer, García-Vargas,
  Garrido, Gesa, Galera, González-Álvarez, González-Cuesta, Grohnert,
  Grözinger, Guàrdia, Guijarro, Hedrosa, Hermann, Hermelo, Arabí, Hernando,
  Hidalgo, Holgado, Huber, Huber, Huke, Kehr, Kim, Klein, Klüter, Klutsch,
  Labarga, Labiche, Lamert, Laun, Lázaro, Lemke, Lenzen, Llamas, Lizon,
  Lodieu, González, López-Morales, Salas, López-Santiago, Madinabeitia,
  Mall, Mancini, Molina, Martínez-Rodríguez, Fernández, Marvin, Mirabet,
  Moreno-Raya, Moya, Mundt, Naranjo, Panduro, Pascual, Pérez-Calpena,
  Perryman, Pluto, Ramón, Redondo, Reinhart, Rhode, Rix, Rodler, Rohloff,
  Sánchez-Blanco, Carrasco, Sarmiento, Schmidt, Storz, Strachan, Stürmer,
  Suárez, Tabernero, Tal-Or, Tulloch, Ulbrich, Veredas, Linares,
  Vidal-Dasilva, Vilardell, Wagner, Winkler, Wolthoff, Xu, \&
  Zhao}]{Quirrenbach_2018}
Quirrenbach, A., Amado, P.~J., Ribas, I., {et~al.} 2018, in Ground-based and
  Airborne Instrumentation for Astronomy VII, ed. C.~J. Evans, L.~Simard, \&
  H.~Takami, Vol. 10702, International Society for Optics and Photonics (SPIE),
  246 -- 263

\bibitem[{Reiners {et~al.}(2018)Reiners, Zechmeister, Caballero, Ribas,
  Morales, Jeffers, Schöfer, Tal-Or, Quirrenbach, Amado, \&
  et~al.}]{Reiners_2018}
Reiners, A., Zechmeister, M., Caballero, J.~A., {et~al.} 2018, Astronomy \&
  Astrophysics, 612, A49

\bibitem[{Ribas {et~al.}(2017)Ribas, Gregg, Boyajian, \& Bolmont}]{Ribas_2017}
Ribas, I., Gregg, M.~D., Boyajian, T.~S., \& Bolmont, E. 2017, Astronomy \&
  Astrophysics, 603, A58

\bibitem[{Sing {et~al.}(2019)Sing, Lavvas, Ballester, Lecavelier~des Etangs,
  Marley, Nikolov, Ben-Jaffel, Bourrier, Buchhave, Deming, \&
  et~al.}]{Sing_2019}
Sing, D.~K., Lavvas, P., Ballester, G.~E., {et~al.} 2019, The Astronomical
  Journal, 158, 91

\bibitem[{Snellen {et~al.}(2010)Snellen, de~Kok, de~Mooij, \&
  Albrecht}]{Snellen_2010}
Snellen, I. A.~G., de~Kok, R.~J., de~Mooij, E. J.~W., \& Albrecht, S. 2010,
  Nature, 465, 1049–1051

\bibitem[{Tennyson {et~al.}(2016)Tennyson, Lodi, McKemmish, \&
  Yurchenko}]{Tennyson_2016a}
Tennyson, J., Lodi, L., McKemmish, L.~K., \& Yurchenko, S.~N. 2016, Journal of
  Physics B: Atomic, Molecular and Optical Physics, 49, 102001

\bibitem[{Tennyson {et~al.}(2020)Tennyson, Yurchenko, Al-Refaie, Clark, Chubb,
  Conway, Dewan, Gorman, Hill, Lynas-Gray, \& et~al.}]{ExoMol_2020}
Tennyson, J., Yurchenko, S.~N., Al-Refaie, A.~F., {et~al.} 2020, Journal of
  Quantitative Spectroscopy and Radiative Transfer, 255, 107228

\bibitem[{{The pandas development team}(2020)}]{Pandas_2020}
{The pandas development team}. 2020, pandas-dev/pandas: Pandas

\bibitem[{van~der Walt {et~al.}(2011)van~der Walt, Colbert, \&
  Varoquaux}]{van_der_Walt_2011}
van~der Walt, S., Colbert, S.~C., \& Varoquaux, G. 2011, Computing in Science
  \& Engineering, 13, 22–30

\bibitem[{Virtanen {et~al.}(2020)Virtanen, Gommers, Oliphant, Haberland, Reddy,
  Cournapeau, Burovski, Peterson, Weckesser, \& et~al.}]{Virtanen_2020}
Virtanen, P., Gommers, R., Oliphant, T.~E., {et~al.} 2020, Nature Methods, 17,
  261–272

\bibitem[{Wilson {et~al.}(2021)Wilson, Gibson, Lothringer, Sing, Mikal-Evans,
  de~Mooij, Nikolov, \& Watson}]{Wilson_2021}
Wilson, J., Gibson, N.~P., Lothringer, J.~D., {et~al.} 2021, Monthly Notices of
  the Royal Astronomical Society, 503, 4787–4801

\bibitem[{Yurchenko {et~al.}(2018)Yurchenko, Al-Refaie, \&
  Tennyson}]{Yurchenko_2018}
Yurchenko, S.~N., Al-Refaie, A.~F., \& Tennyson, J. 2018, Astronomy \&
  Astrophysics, 614, A131

\end{thebibliography}

\begin{appendix}

\onecolumn

\section{Model transmission spectra}
\begin{figure*}[h]
    \centering
    \includegraphics[width=17cm]{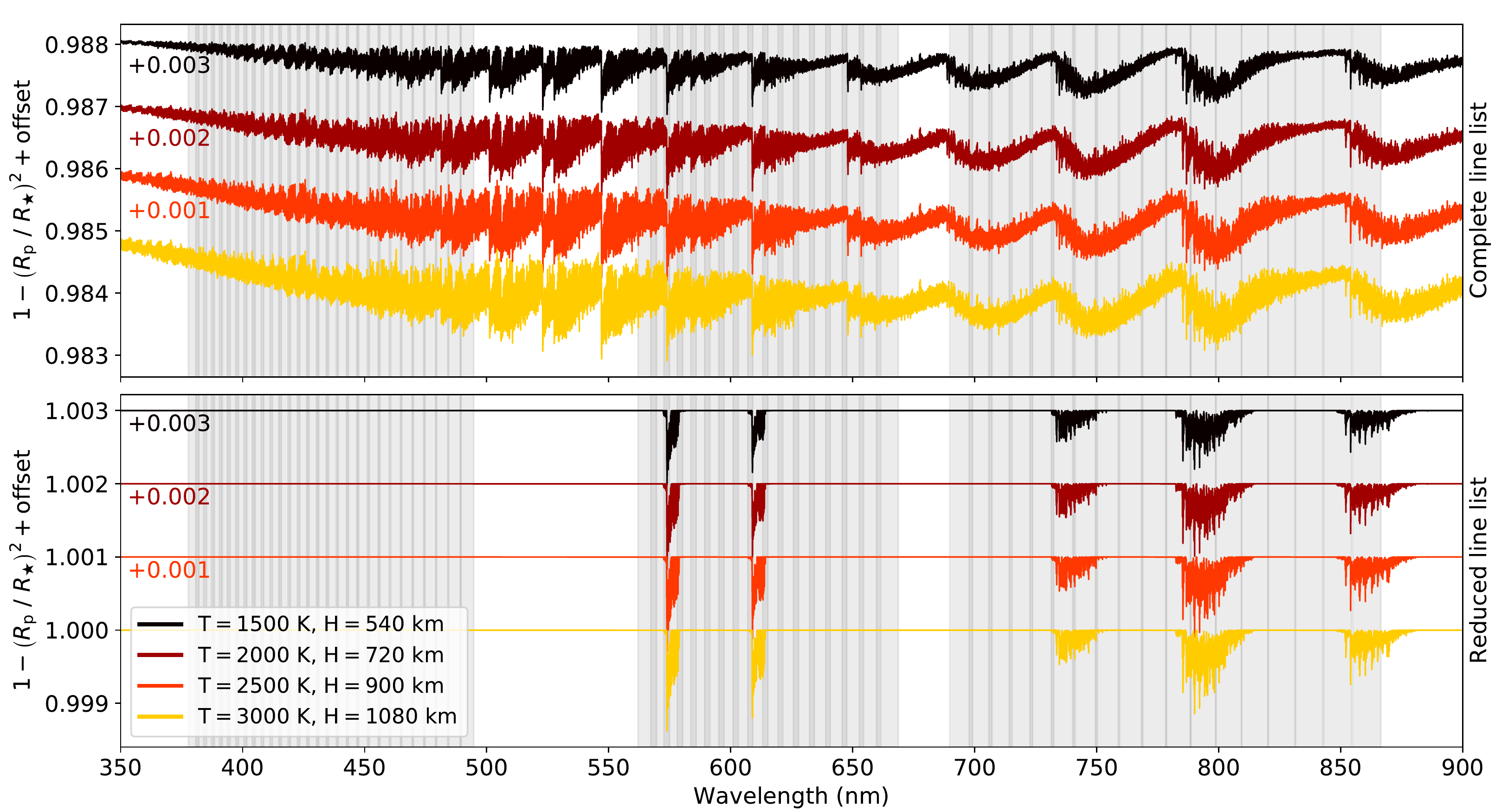}
    \caption{Examples of VO model transmission spectra using an abundance of $\mathrm{VMR}=10^{-6.6}$. The top panel shows the spectra generated with the complete line list and the bottom panel used the pseudo-continuum correction described in Sect. \ref{Section2.4} to create transmission spectra of the reduced line list. The four temperatures used in the presented analysis are shown with different colours and are offset for easier comparison. The grey vertical bands in both panels show the wavelength ranges of the UVES spectral orders.}
    \label{fig:model_transmission_spectra}
\end{figure*}

\end{appendix}

\end{document}